\documentclass[sigconf]{acmart}

\AtBeginDocument{%
  }

\usepackage{graphicx}
\usepackage{subfigure}
\usepackage{makecell}
\providecommand{\ie}{\emph{i.e.,} }
\providecommand{\eg}{\emph{e.g.,} }

\usepackage{tikz}

\newcommand\ChapterPrecis[2]{%
\begin{tikzpicture}[remember picture,overlay]
\node[anchor=north, draw=black, fill=yellow!20, inner sep=5pt, rounded corners, align=left, yshift=-#1] at (current page.north) 
{\parbox[t][1.1cm][l]{16cm}{\small #2}};
\end{tikzpicture}%
}

\newcommand{\mmnote}[1]{\textcolor{black}{#1}}
\copyrightyear{2025}
\acmYear{2025}
\setcopyright{cc}
\acmConference[KDD '25]{Proceedings of the 31st ACM SIGKDD Conference on Knowledge Discovery and Data Mining V.1}{August 3--7, 2025}{Toronto, ON, Canada}
\acmBooktitle{Proceedings of the 31st ACM SIGKDD Conference on Knowledge Discovery and Data Mining V.1 (KDD '25), August 3--7, 2025, Toronto, ON, Canada}
\acmDOI{10.1145/3690624.3709397}
\acmISBN{979-8-4007-1245-6/25/08}

\begin{document}

\title{TGDataset: Collecting and Exploring\\ the Largest Telegram Channels Dataset}

\author{Massimo La Morgia}
\affiliation{%
  \institution{Sapienza University of Rome, Department of Computer Science }
  \city{Rome}
  \country{Italy}
  }
\email{lamorgia@di.uniroma1.it}

\author{Alessandro Mei}
\affiliation{%
  \institution{Sapienza University of Rome, Department of Computer Science }
  \city{Rome}
  \country{Italy}
  }
\email{mei@di.uniroma1.it}

\author{Alberto Maria Mongardini}
\affiliation{%
  \institution{Sapienza University of Rome, Department of Computer Science }
  \city{Rome}
  \country{Italy}
  }
\email{mongardini@di.uniroma1.it}

%\renewcommand{\shortauthors}{La Morgia et al.}
\renewcommand{\shortauthors}{Massimo La Morgia, Alessandro Mei, and Alberto Maria Mongardini}

\begin{abstract}
Telegram is a widely adopted instant messaging platform.
It has become worldwide popular because of its emphasis on privacy and its social network features such as channels---virtual rooms in
which only the admins can post and broadcast messages to all the subscribers.
Channels are used to deliver live updates (\eg weather alerts) and content to a large audience (\eg COVID-19 announcements) but unfortunately also to disseminate radical ideologies and coordinate attacks such as the Capitol Hill riot.

This paper introduces the TGDataset, the most extensive publicly available collection of Telegram channels, comprising 120,979 channels and over 400 million messages. We outline the data collection process and provide a comprehensive overview of the data set. Using language detection, we identify the predominant languages within the dataset. We then focus on English channels, employing topic modeling to analyze the subjects they cover.
Finally, we discuss some use cases in which our dataset can be instrumental in understanding the Telegram ecosystem and studying the diffusion of questionable news.
Alongside the raw dataset, we release the scripts used in our analysis, as well as a list of channels associated with a novel conspiracy theory known as Sabmyk.
\end{abstract}



\keywords{Dataset, Telegram, Conspiracy Theories, Copyright Infringement}



\maketitle

%%%%% Arxiv Reference
    \ChapterPrecis{1.0cm}{If you cite this paper, please use the ACM SIGKDD Conference on Knowledge Discovery and Data Mining reference: Massimo La Morgia, Alessandro Mei, Alberto Maria Mongardini. 2025. TGDataset: Collecting and Exploring the Largest Telegram Channels Dataset. \textit{KDD.} '25, August 3–7, 2025, Toronto, ON, Canada, 11 pages. \color{blue}{\url{https://doi.org/10.1145/3690624.3709397}}
}
%%%%

\section{Introduction}
In today's digital age, instant messaging apps have become ubiquitous, and Telegram stands out as one of the leading platforms. Its focus on user privacy is a significant factor behind its growing popularity~\cite{fortune}.
However, as is often the case with platforms that prioritize user privacy, this approach has also attracted malicious users who exploit the platform for illegal activities.
While Telegram serves legitimate purposes, such as disseminating weather alerts or government updates~\cite{covidTelegram}, it is also misused for spreading radical ideologies~\cite{telegramNeoNazi}, orchestrating violent attacks~\cite{bbcterrorist}, and engaging in market manipulations, such as pump-and-dump schemes~\cite{la2020pump}. These dual uses underscore Telegram's complex nature and highlight the need for a deeper exploration of its ecosystem.

In this study, we introduce and publicly release~\cite{TGDatasetRepo} the TGDataset, a collection of public Telegram channels of over 120,000 channels and 400 million messages. 
To the best of our knowledge, the TGDataset surpasses existing datasets in scale and scope, offering an unprecedented opportunity to study the Telegram ecosystem comprehensively. Unlike prior datasets focused on specific topics or geographic regions, the TGDataset includes diverse channels covering multiple languages and subjects, enabling a more holistic understanding of the platform.
\mmnote{
In particular, the TGDataset enables the study of the platform's political leanings and the sharing of questionable content, topics widely explored on other Online Social Networks (OSNs) but not on Telegram. It also includes numerous channels spreading conspiracy theories, as well as those engaged in borderline or illegal activities like carding, inciting violence, and promoting Nazi ideologies. Thus, with its scale and diversity, the TGDataset represents a valuable resource for researchers willing to investigate and shed light on both the legitimate and problematic aspects of Telegram’s ecosystem.} 

\mmnote{
The main contributions of this work are the following:
\begin{itemize}
    \item We present the largest publicly available dataset of Telegram channels, enabling researchers to explore various phenomena on the platform. We perform analyses such as language detection and topic modeling to characterize the dataset’s contents, revealing dominant languages, prevalent topics, and temporal trends in channel creation.
    \item We explore several potential use cases of the TGDataset. One application involves analyzing the diffusion of questionable content and biases across Telegram channels, providing insights into content moderation and information reliability. Another use case focuses on investigating networks that promote conspiracy theories, including emerging phenomena such as the Sabmyk network. Additionally, the dataset can be used to investigate channels engaged in borderline or illegal activities, such as carding, copyright infringement, and extremist propaganda.
    \item Alongside the dataset, we release scripts for analysis, language detection, and topic modeling, fostering reproducibility and facilitating further research.
\end{itemize}
}

\begin{figure*}
  \Description{A figure showing three screenshots of a Telegram channel interface: (A) displays the channel's header with four labeled elements - the title, subscriber count, description, and username; (B) shows the feed of posts published by the channel; (C) shows an example of a forwarded message within the channel.}
  \includegraphics[width=.75\textwidth]{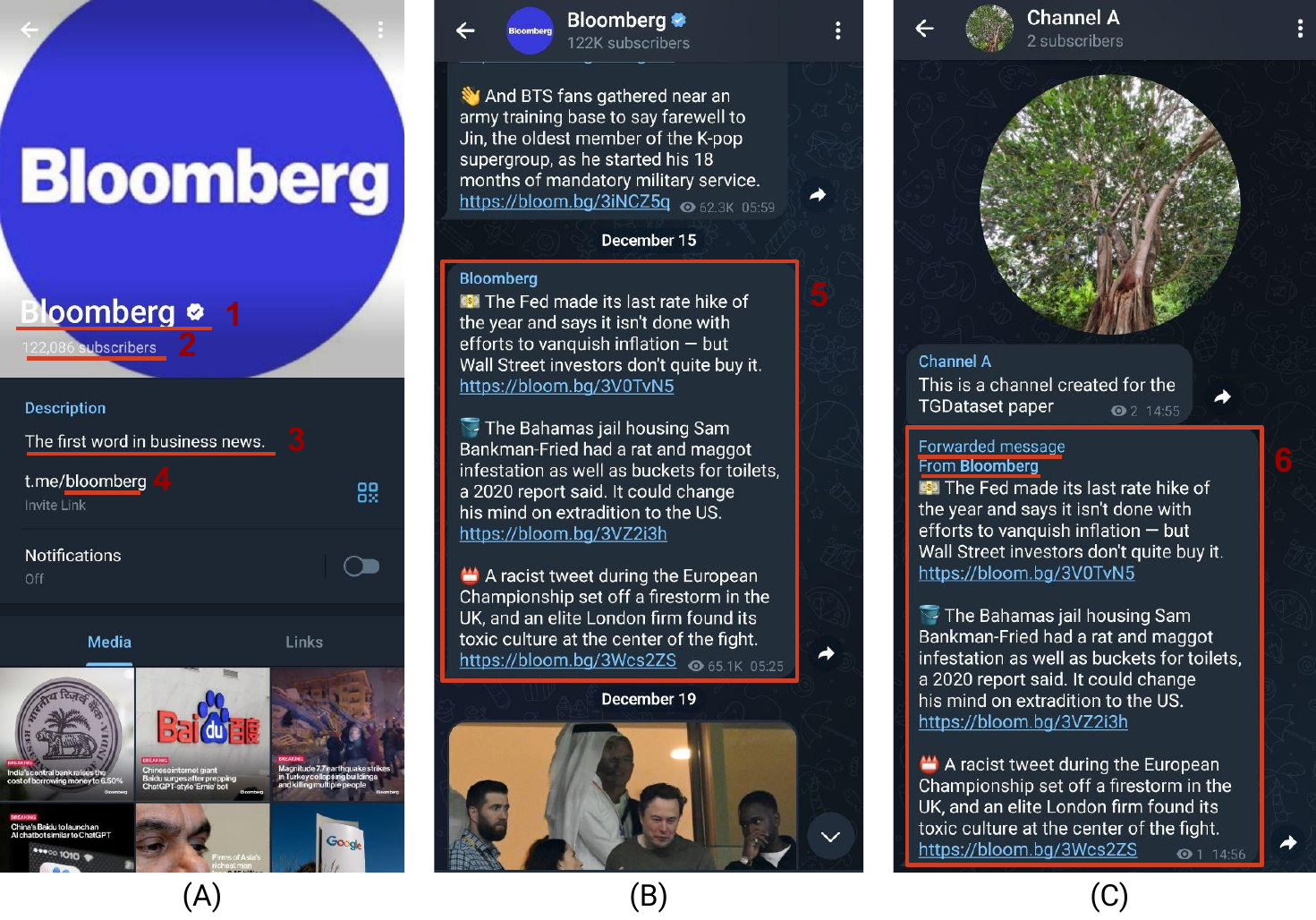}
  \caption{Fig.~(A) represents the main information about a channel: the title (1), the number of subscribers (2), the description (3), and the username (4). Fig.~(B) shows the posts published by the channel (5). Fig.~(C) displays a forwarded message (6).   
  }
  \label{fig:screenshots_example}
\end{figure*}

\section{Background: Telegram}
Started in 2013, Telegram is now a popular instant messaging platform, with more than 500 million active users in 2021~\cite{TelegramUsers}. 
Telegram allows users to post content such as text messages, images, audio, stickers, videos,  and files weighing up to 2~GB.
In addition to private chats (one-to-one communication), Telegram provides two other ways to communicate: Through groups and channels.

\textbf{Groups}: Designed for many-to-many messaging, groups allow up to 200,000 members to communicate with each other by posting content. Groups are identified by a unique username and have a description and a title. They can be public (\ie can be found with Telegram search, and every user can read its content and join the group) or private (\ie only members can see the posts, and users need an invite link to join it). Besides, a group presents a list of its members which, if the group is public, is visible to anyone (even non-members). 

\textbf{Channels}: They offer one-to-many communication. Thus, differently from groups where every member can post messages, in channels, only administrators can publish posts. Channels have a unique username, a title, a description, and can be private or public. Fig.~\ref{fig:screenshots_example} presents the information displayed by a channel.
Contrary to groups, a channel can have an unlimited number of subscribers and does not display any users' information (\ie nicknames) other than the number of its subscribers.
Thanks to these features, Telegram channels have become popular for disseminating content to a large audience. Indeed, several public figures and companies started using official Telegram channels to share news and updates~\cite{covidTelegram}. 
Nonetheless, this tendency caused an increase in the number of channels on the platform seeking to impersonate official channels or sell counterfeit products and services.

\textbf{Scam and Verified mark}: To tackle this issue, Telegram has implemented the \textit{verified} and the \textit{scam} marks. The verified mark requires that channels, groups, and bots prove their verified status on at least two social media platforms(\eg TikTok, Facebook, Twitter, Instagram)~\cite{howVeryfyTelegramCh}. In contrast, the scam status is assigned to channels and groups reported for fraud by multiple users~\cite{telegramScam}.

\textbf{The forward mechanism}: Although private chats, groups, and channels can not communicate directly among them, they are not isolated but may be linked through message forwarding. Leveraging this feature, users and channels can forward messages to other users, groups, or channels displaying the post's original author. 
Fig.~\ref{fig:screenshots_example} shows an example of message forwarding, where the admin of "Channel A" (Fig.~\ref{fig:screenshots_example}~(C)) forwards in his channel a message from the Bloomberg's channel. The message in the red square of Fig.~\ref{fig:screenshots_example}~(B)) is the original message, while the one in the red square of Fig.~\ref{fig:screenshots_example}~(C) is the forwarded message. As shown in Fig.~\ref{fig:screenshots_example}, the forwarded message displays the name of the original channel of the post, which, if clicked, leads to that channel.

\section{Related works}
Several studies have been conducted on the Telegram platform and related issues. Hashemi et al.~\cite{hashemi2019telegram} collected Telegram Iranian channels and groups to distinguish high-quality groups (\eg business groups) from low-quality groups (\eg dating groups). They found that high-quality groups had longer messages and more user engagement. Jalilvand et al.~\cite{jalilvand2020channel} tackled the challenge of ranking channels related to a user request on Telegram. Ng et al.~\cite{ng2020pofma} analyzed a Singapore-based COVID-19 Telegram group with over 10,000 participants, examining group opinions over time. Their work revealed a peak in engagement when the Ministry of Health raised the disease alert level but a subsequent decline in participation.
Nobari et al.~\cite{dargahi2017analysis} performed a structural and topical analysis of Telegram messages, building a dataset of over 2,500 channels and 54 groups and constructing a graph based on mentions. This study found that the PageRank algorithm was not effective in identifying high-quality Telegram channels. \mmnote{Analyzing the dataset released by Nobari et al., the TGDataset contains more than ten times the number of channels and is significantly more recent (July 2022 versus October 2016).}

Similar to our work, Baumgartner et al.~\cite{baumgartner2020pushshift} created the Pushshift Telegram dataset, containing both channels and groups. They collected over 27,800 channels and 317 million messages from 2.2 million unique users updated to November 2019. %, including right-wing extremist groups and protest movements. 
%27,944 canali
%1,885 gruppi
%317,224,716 messaggi:
    %192,044,689 messaggi dei canali
    %125,180,027 messaggi dei gruppi
Differently from~\cite{baumgartner2020pushshift}, the TGDataset boasts a more recent dataset, with the message history of channels aligned to July 2022, thus almost three years of more data. 
Moreover, while the Pushshift dataset contains both channels and groups, the TGDaset is made of only channels. 
Thus, narrowing the comparison on the Telegram channels, the TGDataset includes a significantly higher number of channels (approximately four times more) and a greater volume of messages posted in channels (more than double).
\mmnote{Table \ref{tab:datasets_comparison} provides the exact number of channels and messages contained in the dataset released by~\cite{dargahi2017analysis}, Pushshift, and TGDataset.}
\begin{table}
    \centering
    \small
    \caption{Summary of the number of channels, channel messages, and group messages present in both the TGDataset and Pushshift.}
    \label{tab:datasets_comparison}
    \begin{tabular}{l|ccc}
    \toprule
        Dataset & \# channels & \# channels msgs & Update to\\ %\# groups msgs\\ 
        \midrule
        Nobari et al.~\cite{dargahi2017analysis} & 2,556 & 36,928   & October 2016 \\ %& 54 groups
        Pushshift~\cite{baumgartner2020pushshift} & 27,801 & 192,044,689    & November 2019 \\ %& 1,885 groups 
        TGDataset & 120,979 & 498,320,597   & July 2022\\ %& 0 groups
        \bottomrule
    \end{tabular}
\end{table}
Finally, another important difference between the two datasets is about the seed choice. Indeed, the Pushshift dataset used as seed channels only those of right-wing extremist politics or cryptocurrencies-related channels, while our dataset starts the collection from seed channels covering heterogeneous topics (see Sec.~\ref{sec:dataset_construction}). This choice led to building a dataset that should better represent the status of the Telegram channels ecosystem.

%\amnote{Note that, among the 317 million messages, the authors included 125 million messages posted in Telegram groups. Therefore, the actual number of messages posted in channels within their dataset is about 192 million messages.
%--- Oppure ---
%Of the 317 million messages, 125 million were posted in Telegram groups. Consequently, the number of messages specifically posted in channels within this dataset amounts to approximately 192 million.
%}
Weerasinghe et al.~\cite{weerasinghe2020pod} found that Telegram hosts organized groups called "pods" where members boost each other's Instagram account popularity through interaction. Other studies have documented the presence of Telegram channels and groups focused on cryptocurrency pump and dump scams~\cite{xu2019anatomy, la2020pump}.
The work of Urman et al.~\cite{urman2022they} explored far-right networks on Telegram, revealing that the most prevalent communities are those related to 4chan~\cite{4chan} and Donald Trump's supporters. Moreover, the authors pointed out that the sudden growth of these networks on Telegram corresponds with the bans of numerous far-right figures on other online social networks.
%Another paper~\cite{hoseini2020demystifying} characterized public groups of different messaging platforms (Whatsapp, Telegram, Discord). They found the latter platform is suitable for discovering public groups of these messaging services. Furthermore, the study showed that group URLs are temporary and that the messaging platforms examined display sensitive user information.
Additionally, numerous studies have investigated the use of Telegram by terrorist organizations such as ISIS for content dissemination and proselytism~\cite{cao2017dynamical, yayla2017telegram}.

\section{Data collection}
\label{sec:data_collection}
Previous Telegram datasets have been created with specific research goals in mind, resulting in a limited scope of channels related to a particular language, country, or topic. For example, \cite{hashemi2019telegram} only includes Iranian channels, ~\cite{hoseini2021globalization} only contains 151 channels related to QAnon, and the PushShift dataset focuses on channels discussing right-wing extremism and cryptocurrency. %Instead, the PushShift dataset~\cite{baumgartner2020pushshift}, to the best of our knowledge, the only other public dataset, contains 27,800 channels focusing on channels discussing right-wing extremism and cryptocurrency. 
In contrast, 
our dataset aims to provide a global snapshot of the Telegram ecosystem. To achieve this goal, we require a dataset that provides an up-to-date representation of Telegram and encompasses a broad range of popular and interconnected channels. This is the motivation behind the creation of the TGDataset.

\subsection{Dataset construction}
\label{sec:dataset_construction}

\begin{figure}
  \Description{A flow diagram illustrating the data collection process. The diagram shows the sequential steps involved in collecting, processing, and storing data from Telegram channels.}
  \includegraphics[width=.44\textwidth]{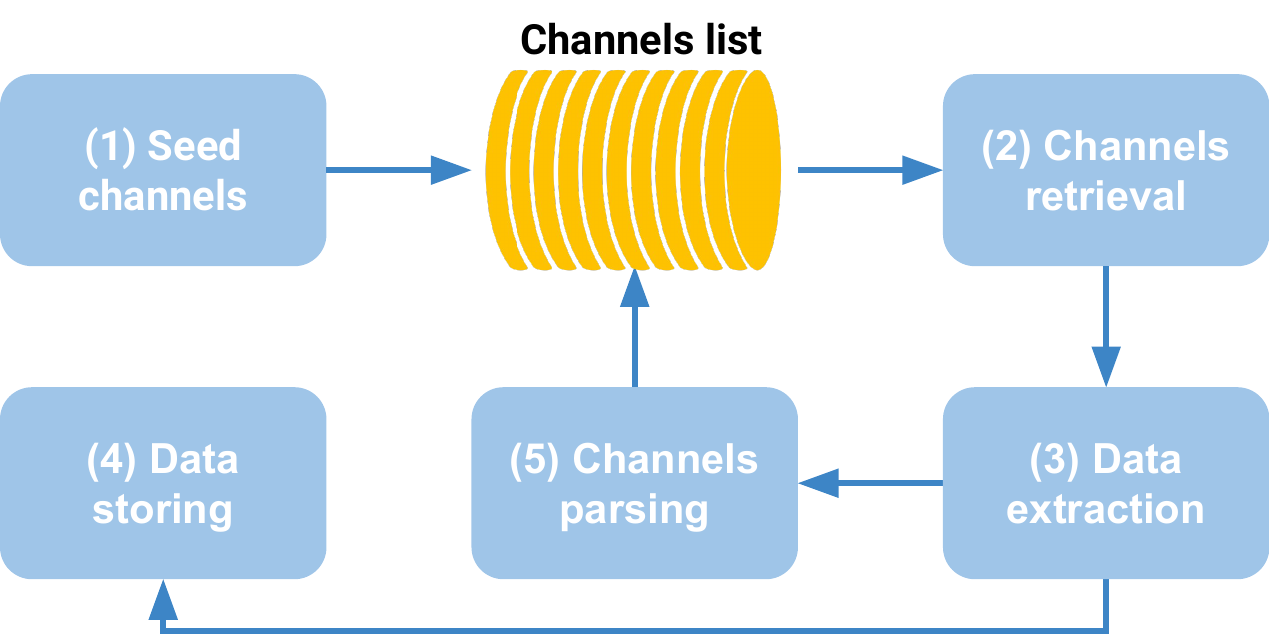}
  \caption{Data collection flow diagram.  
  }
  \label{fig:data_collection_flow}
\end{figure}

%To examine Telegram, specifically the most well-known and interconnected channels, we begin with a set of seed channels related to various topics. We then extend our dataset by including the source channel for every forwarded message found in the seed channels, as previously outlined in
To build the TGDataset, we use a snowball approach as previously done in~\cite{baumgartner2020pushshift}. We begin with a set of seed channels related to various topics and then extend our dataset by including the source channel of each forwarded message.
Fig.~\ref{fig:data_collection_flow} represents the flow diagram we use to collect the data. 
To determine the seed channels (Step 1 in Fig.~\ref{fig:data_collection_flow}), we utilize Tgstat~\cite{tgstat}, a popular freemium service that indexes over 150,000 Telegram channels and gathers statistics about them. Tgstat provides information for each channel, such as the subscriber count, topic category, growth rate, and language. Among the other metrics, Tgstat lists the top 100 channels by the number of subscribers. From this rank, we retrieve all the categories these channels belong to, finding the following 18 categories:: \textit{Sales}, \textit{Humor and entertainment}, \textit{News and Mass media}, \textit{Video \& Movies}, \textit{Business \& Startups}, \textit{Cryptocurrencies}, \textit{Politics}, \textit{Technologies}, \textit{Sport}, \textit{Marketing}, \textit{Economics}, \textit{Games}, \textit{Religion}, \textit{Software and Applications}, \textit{Lifehacks}, \textit{Fashion \& Beauty}, \textit{Medicine}, and \textit{Adults}. Finally, we selected the ten most popular channels from each category as the seed channels according to their subscriber count.
With this approach, we collect 180 seed channels.

For each seed channel, we proceed with downloading all its messages by leveraging the Telethon APIs~\cite{telethonAPI} (Step (2)), an open-source Python tool that provides access to the official Telegram APIs. Specifically, each call to download messages from a channel is performed five seconds after the previous one to avoid flooding the Telegram service.  
Once we retrieved the data about the channel, we proceeded with the data extraction (Step (3)): in particular, we retained all the information obtained by Telegram API, excluding media files such as images, videos, or PDFs. The reason behind this decision is to avoid storing copyrighted or illegal content. Anyway, even for media-based messages, we collect the name, author, and ID of the media content using the Telegram API, and we infer the file format based on the extension of the file name. 
\mmnote{Next, we store the extracted information in MongoDB~\cite{mongodb}, a NoSQL database (Step (4)).} Finally, we parse the channel messages to identify forwarded messages and their original authors (Step (5)).
If the author represents a channel that we have never seen, then the channel is added to the list of new channels. Conversely, if the author is a user or represents a group, we ignore the author of the forwarded message.  
To iteratively expand the TGDataset, we repeat the above procedure of channel discovery (Steps 2, 3, 4, and 5) using the newly found channels as new seeds. %After three iterations, more than half of the seed channels no longer contributed to the dataset as all their forwarded messages had been fully explored. 
We stopped the iteration process on July 31, 2022.
Finally, to align our dataset, for each discovered channel, we download all the missing messages until July 31, 2022. % the then we align all the channel's messages. 

For each channel, we store the following information as listed in Tab.~\ref{tab:data_collected}: title and if it is marked as verified or scam (item~1 in Fig.~\ref{fig:screenshots_example}~(A)), subscriber count (item~2), description (item~3), username (item~4), ID, and creation date. For the messages, we retain the author (the name of the channel), timestamp, and, in the case of forwarded messages, original posting date and original channel.  
%Finally, we store the text content of text-based messages and just the title and file format of media-based messages.

\begin{table}[]
    \centering
    \small
    \begin{tabular}{ll}
    \toprule
        \textbf{Field name}  & \textbf{Description}\\
        \midrule
        Channel ID & the ID of Telegram channel \\
        Creation date & the timestamp of the creation date of the channel \\
        Username & the Telegram username of the channel \\
        Title & the title of the channel \\
        Description & the description of the channel \\
        Scam & indicates if Telegram marked the channel as a scam \\
        Verified & indicates if Telegram marked the channel as verified \\
        \# subscribers & the number of subscribers of the channel \\
        \hline
        Text messages & \\%the text messages posted in the channel\\
        \hline
        Message & the text of the message\\
        Date & the timestamp of the message publication\\
        Author & the ID of who posted the message\\
        Is forwarded & indicates if the message is forwarded\\
        Forwarded from & the ID from which the message is forwarded\\
        Original date & the timestamp of the first post of the message\\
        \hline
        Media messages & \\
        \hline
        Title & the title of the content\\
        Media ID & the ID of the content on Telegram\\
        Date & the timestamp of the content\\
        Author & the ID of who posted the content (int)\\
        Extension & the format of the content (string)\\
        Is forwarded & indicates if the content is forwarded\\
        Forwarded from & the ID from which the content is forwarded\\
        Original date & the timestamp of the first post of the content\\
        \bottomrule
    \end{tabular}
    \caption{Information stored about collected channels}
    \label{tab:data_collected} 
\end{table}

\subsection{Accessing the TGDataset and FAIR principles}
We released the TGDataset in alignment with the FAIR (Findable, Accessible, Interoperable, Reusable) principles~\cite{wilkinson2016fair}.

\textbf{Findable.} We publicly released the        TGDataset~\cite{TGDataset} on Zenodo~\cite{zenodo}, an open repository managed by CERN.

\textbf{Accessible.} The dataset is made of 121 files, each of which contains a maximum of 1,000 channels. To ease the retrieval of single channels from the dataset, we store them in alphabetical order. In addition, each file's name describes the index of the channels it contains.
Given the large dimension of the dataset, even if compressed (approx. 71GB), we divided the dataset into four parts. In this way, it is possible to download and explore also a single portion of the dataset. Moreover, we released a sample of the dataset on a public GitHub repository~\cite{TGDatasetRepo}. This enables users to explore the dataset's structure without the need to download any portion of it.

\textbf{Interoperable.} All the information is encoded in JSON format so that the data can be easily parsed and manipulated with most programming languages.

%We also released, on a public GitHub repository~\cite{TGDatasetRepo}, Python scripts to load the dataset into a MongoDB~\cite{banker2016mongodb} database, the  scripts that illustrate how to perform basic queries to the database (\eg insertion of new channels, retrieve channels by name or by id), and to replicate some of the analysis we perform in this work, such as the Language Detection of the channels~(Sec.~\ref{sec:language_detection}) and the Topic Modeling~(Sec.~\ref{sec:topic_modeling}).
\textbf{Reusable.} We have outlined the methodology employed for data collection, and the data is open for anyone to use.
We also released on the public GitHub repository~\cite{TGDatasetRepo}, Python scripts to load the data into a MongoDB~\cite{banker2016mongodb} database, as well as scripts that show how to perform basic queries on the database (such as inserting new channels or retrieving channels by username or ID). 
Furthermore, these scripts can also replicate some of the analysis that we performed in this work, including Language Detection of the channels~(Section~\ref{sec:language_detection}) and Topic Modeling~(Section~\ref{sec:topic_modeling}).
Finally, we released as CSV files the list of channels belonging to the Sabmyk network and the list of other channels promoting conspiracy theories (Sec.~\ref{sec:sabmyk}), the list of channels annotated with the detected writing language, and the one containing the mapping of inferred topics.

\subsection{Dataset overview}
The data collection for the TGDataset began on 4 January 2021 and ended on 31 July 2022. The dataset has a total size of about 460 GB and includes 498,320,597 messages and 120,979 unique channels. Out of these channels, 654 (0.54\%) have verified status, while 183 (0.15\%) are flagged as scam channels. %Upon manual examination, we found that these scam channels mainly relate to trading, cryptocurrencies, and fake accounts of political figures such as Donald Trump, Mike Pompeo, and Ivanka Trump.
For the purpose of clarity, we will refer to the remaining 120,142 channels, which are neither scams nor verified, as \textit{standard channels}.

\begin{figure*}[h!]
  \centering
  \Description{A figure containing three CDF plots comparing scam, verified, and standard Telegram channels. The first plot shows the distribution of subscriber counts, the second shows the distribution of total messages posted, and the third shows the distribution of forwarded message ratios. These plots help visualize the differences in behavior and characteristics between the different types of channels.}
  \subfigure[Subscribers]{%
  \includegraphics[width=.32\textwidth]{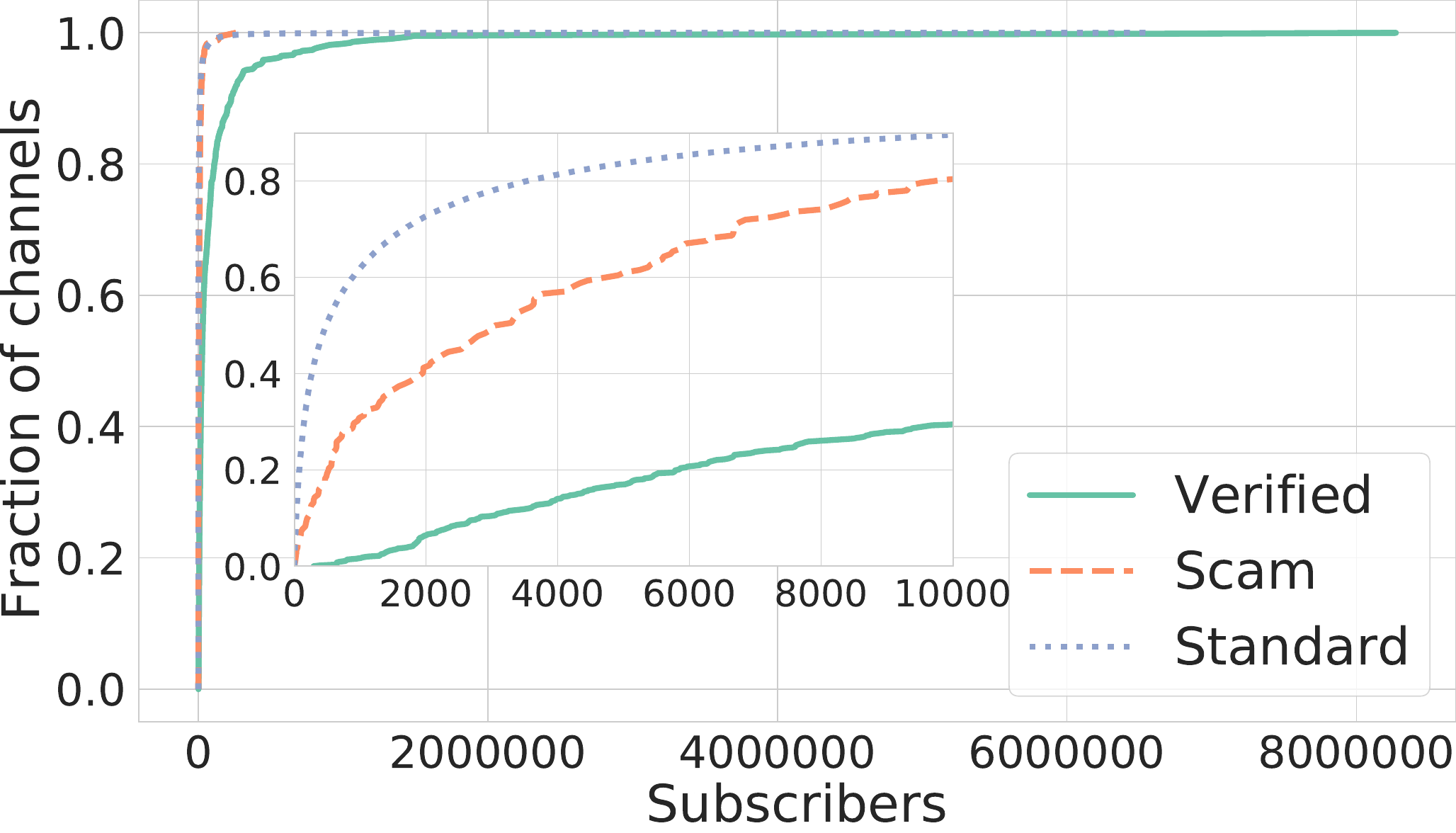}
  \label{fig:cdf_subscribers}}
  \subfigure[Messages.]{%
  \includegraphics[width=.32\textwidth]{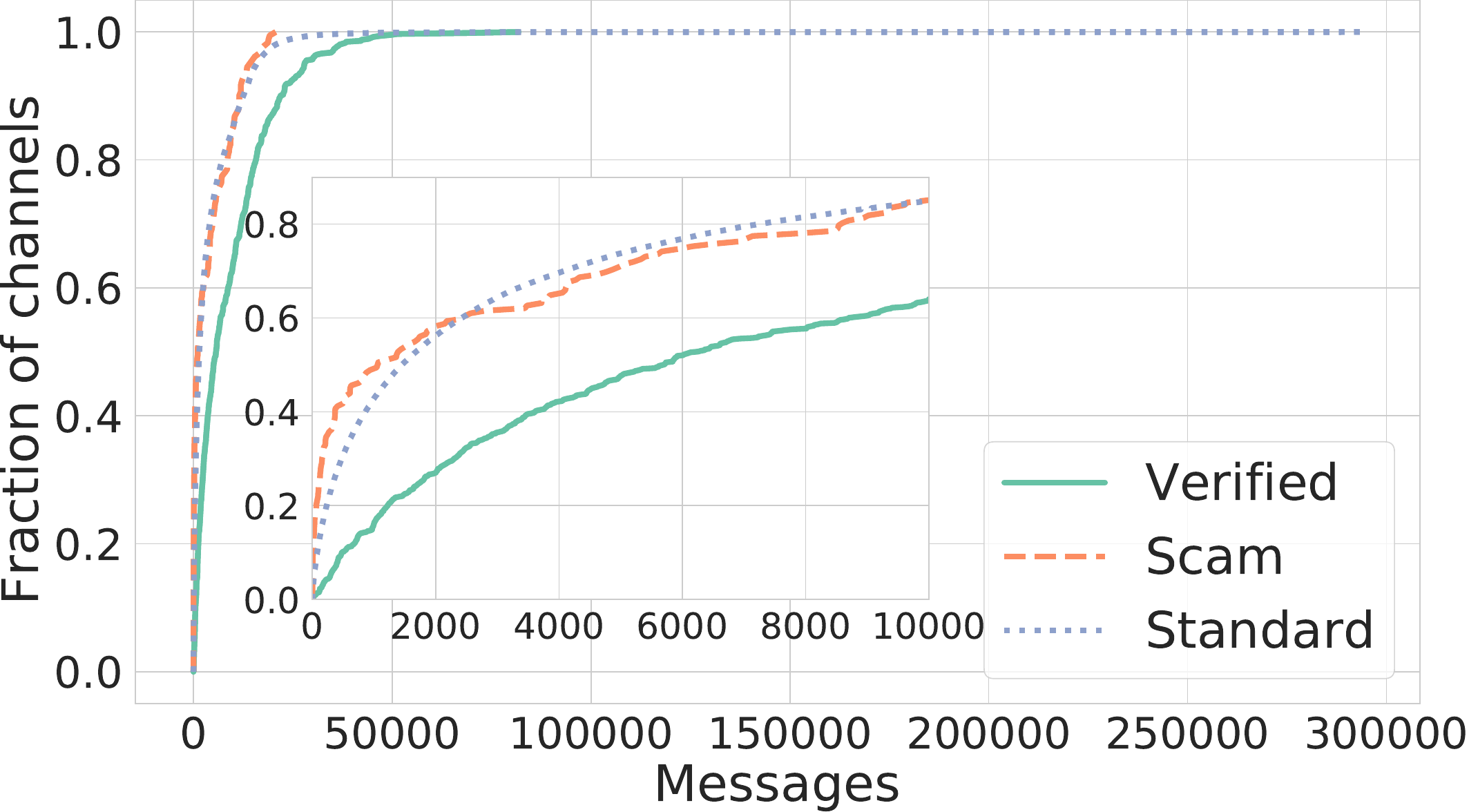}
  \label{fig:cdf_messages}}
  \subfigure[Forwarded messages]{%
  \includegraphics[width=.32\textwidth]{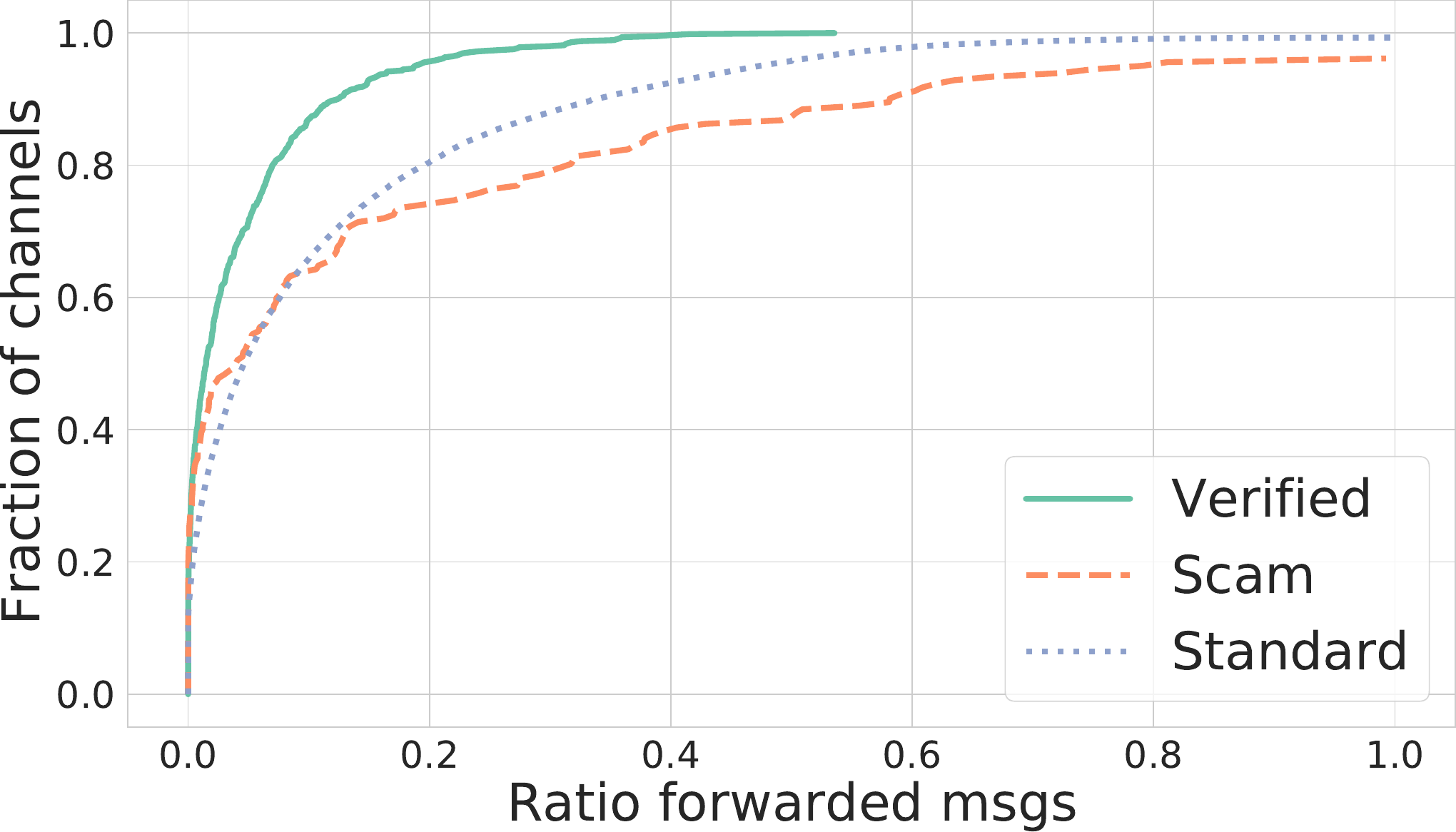}
  \label{fig:cdf_forwarded_messages}}
  \caption{CDFs of the number of subscribers for scam, verified, and standard channels (\ref{fig:cdf_subscribers}), 
  the number of messages posted (\ref{fig:cdf_messages}), and ratio of forwarded messages (\ref{fig:cdf_forwarded_messages}).} 
  \label{fig:cdf_dataset}
\end{figure*}

To provide a general characterization of the channels contained in the TGDataset, in Fig.~\ref{fig:cdf_dataset}, we show the CDFs of the number of subscribers, posts published, and the ratio of forwarded messages for verified, scam, and standard channels. 
As shown in Fig.~\ref{fig:cdf_subscribers}, verified channels (green line), in general, have more subscribers than scam (dashed orange line) and standard channels (dotted blue line). Indeed, the verified channels with more than 1,000 subscribers are 98.32\%, while the scams and standard are 68.85\% and 38.11\%, respectively. 
%equal or greater than 10,000
%standard -  0.1036
%scams -  0.1967
%verified -  0.7049
Concerning the number of messages published, Fig.~\ref{fig:cdf_messages} suggests that verified channels tend to post more content than scams and standard ones. The standard channels with at least
1,000 messages are 57.16\%, those verified are 83.79\%, and the scams
are 50.82\%.
%equal or greater than 10,000
%standard -  0.1497
%scams -  0.153
%verified -  0.3609
Moreover, we report the extent of use of the forwarding function among channels of the TGDataset. Fig.~\ref{fig:cdf_forwarded_messages} reveals message forwarding is more prevalent in standard and scam channels. The standard channels with at least a 0.1 ratio of their messages that are forwarded are 33.5\%, whereas the scams and the verified ones are 32.24\% and 13.15\%, respectively. 

%\mlnote{Finally, the CDFs in Fig.~\ref{fig:cdf_dataset} (3a and 3b?)illustrate the long tail phenomenon in our dataset, revealing that most channels have relatively few subscribers and posts, while a small number of channels, forming the "head," account for a disproportionately large share of these metrics. This distribution is characteristic of the long tail phenomenon, where the aggregate impact of the numerous smaller entities ("tail") rivals or even exceeds that of the few dominant ones ("head"). 
%Interestingly, this behavior is consistent across standard, scam, and verified channels, indicating that the phenomenon is intrinsic to the broader channel ecosystem rather than specific to a particular channel type.}
\section{Dataset analysis}
In this section, we first study the languages spoken within our dataset, leveraging language detection, and explore the English channels to understand the covered topics using topic modeling. Then, we analyze the temporal aspects of the TGDataset.% focusing on the evolution of Telegram over time and according to its bans in countries or changes of policies in other platforms. Lastly, we build the graph representing the TGDataset and return some metrics about it. 

\subsection{Language detection}
\label{sec:language_detection}
As a first step in our analysis, we want to understand the language coverage of our dataset. This study gives hints about the popularity of the Telegram platform in different regions of the world.
To automatically retrieve the language used in a channel, we leverage Language Detection~\cite{nakatani2010langdetect}, a language detection library implemented in Java by Google with over $99\%$ of precision for $53$ languages.
To get meaningful results dealing with text-based messages, first, the text has to be normalized and polished from useless information~\cite{uysal2014impact}. Hence, we applied the following preprocessing steps. 
First, we excluded messages shorter than 15 characters, as they do not provide sufficient information for accurate analysis~\cite{baldwin2010language}. Next, we utilized the RegexpTokenizer developed by NLTK~\cite{tokenizer} to split text-based messages into tokens. Then, we removed all numbers and emojis from the documents, as they do not contribute to language identification. Also, we discarded references and mentions of groups, channels, or users, as usernames typically do not aid in identifying the language used.

We assign a channel to a specific language only if over 50\% of its messages are written in that language. Otherwise, we will not associate a language with that channel.
At the end of the analysis, we find that the five most spoken languages in our dataset are Russian with 42,983 channels (35.52\%), English with 19,768 channels (16.34\%), Farsi with 16,779 channels (13.86\%), German 4,950 (4.09\%), and Arabic with 2,523 (2.08\%). 
Of particular interest is the strong presence of the Russian and Farsi channels, which likely reflects the popularity of Telegram in Russia and Iran, despite Russia's government banned Telegram from 2018 to 2020~\cite{liftRussianban}, and it is still banned in Iran~\cite{akbari2019platform}.
Although Iran banned Telegram on 30 April 2018~\cite{IranBanTelegram}, in our dataset, there are 4,635 out of 16,779 (27.62\%) channels in Farsi created after the date of the ban. Instead, looking at the last activity of Farsi's channels, we find that more than 66.53\% (11,164) channels continued to operate after the ban.
Regarding Russian channels, as shown in section Sec.~\ref{sec:temporal}, they continue to appear also after the ban of 2018.

\subsection{Topic modeling}
\label{sec:topic_modeling}

In this subsection, we investigate the topics covered by the channels in our TGDataset using Topic Modeling~\cite{hofmann2001unsupervised}.
This is a data mining tool that allows finding a brief description of the topic addressed by the messages of a channel.
For this analysis, we consider only channels that post English content. Thus, we use as input to the topic modeling 19,768 channels (16.34\% of our dataset).
As in language detection, a preprocessing phase is necessary. In addition to the deletion of numbers, deletion of emoji, and normalization, we perform other preprocessing steps.
We exclude words with one or two characters, as they are common in messages (e.g., y, no, ok) and do not provide meaningful information about the topic. We use the \textit{en\_core\_web\_sm} model developed by spaCy~\cite{spacy2} to reduce inflected words to their lemmas, ensuring that different forms of a word are treated as a single term. Lastly, we employ the English stopwords list provided by NLTK~\cite{stopwords} to discard common words (e.g., articles and prepositions) that do not carry significant meaning in a sentence~\cite{schutze1998automatic}.

%As for the language detection task, even in this case, a preprocessing phase is necessary. In addition to the preprocessing operations carried out in the previous task, in this case, we also apply a lowercase normalization and remove single character words (\eg $k$, $y$, $n$) since these words are very commons in the messages and do not provide information about the topic. Finally, we reduce inflected words to their lemma to consider the different inflections of the same word as a unique one (Lemmatization).
To discover the latent topics addressed within the channels, we use the Latent Dirichlet Allocation (LDA)~\cite{blei2003latent} as the Topic Modeling algorithm. LDA needs as input the number of topics, so we used the UMass measure~\cite{mimno2011optimizing} to select the optimal one.
%LDA relies on the idea that documents are generated by a particular probabilistic model, according to which each document is composed of a mixture of a small number of topics, and each word belongs to one of them. 
%UMass is an intrinsic measure of Topic Coherence~\cite{stevens2012exploring}. It computes the log-likelihood that two words representing the topic occur in the same documents. 
In particular, the higher the coherence of the words representing topics, the closer to 0 the value of UMass. In our case, we calculate the best UMass value reached by selecting the number of topics from 10 to 30.

\begin{table}[h!]
    \centering
    \small
    \begin{tabular}{ll}
    \toprule
        Topic & Top 10 keywords\\
        \midrule
        \makecell[vl]{Video-game\\ modding} & \makecell[vl]{pubg, esp, lvl, max, mod, login,\\ password, aimbot, vip, recoil}\\
        \hline
        Carding & \makecell[vl]{iphone, premium, netflix, amazon, samsung,\\ paytm, giveaway, hacking, tutorial, vpn} \\
        \hline
        Entertainment & \makecell[vl]{reddit, artist, submit, edition, meta,\\ til, hire, score, anime, league}\\
        \hline
        Indian education & \makecell[vl]{indian, upsc, exam, affair, copyright,\\ infringement, unavailable, wildlife, prelim, batch} \\
        \hline
        Religion & \makecell[vl]{jesus, lord, christ, spirit, pray, holy, shall,\\ church, prayer, bless}\\
        \hline
        US news & \makecell[vl]{trump, biden, vaccine, election, covid, joe,\\ donald, court, military, mandate}\\
        \hline
        Social & \makecell[vl]{twitter, reuter, tweet, album, telegram, ivurl,\\ utc, radio, hashtag, youtube}\\
        \hline
        World News  & \makecell[vl]{russia, ukraine, independent, minister, \\military, coronavirus, europe, foreign, israel, court}\\
        \hline
        Software & \makecell[vl]{android, rupee, enroll, web, udemy, proxy, \\linux, software, premium, mod}\\
        \hline
        COVID-19 & \makecell[vl]{vaccine, covid19, covid, medical, vaccination, \\vaccinate, pfizer, disease, patient, australia}\\
        \hline
        Crypto & \makecell[vl]{btc, bitcoin, crypto, usdt, binance, usd, token,\\ trading, blockchain, profit}\\
        \hline
        \makecell[vl]{Extremists \\and radicals} & \makecell[vl]{violate, defense, jews, jewish,\\ attacker, defend, jew, defender, hitler, antifa}\\
        \hline
        Adult content & \makecell[vl]{pornographic, leak, t.me, xxx, meme, wanna,\\ sharp, iphone, porn, teen}\\
    \bottomrule
    \end{tabular}
    \caption{Top 10 keywords within LDA topics.}
    \label{tab:LDA_topics} 
\end{table}
The best UMass value (-0.69) is obtained with 13 topics. Tab.~\ref{tab:LDA_topics} shows the inferred topics and the top 10 keywords for each of them.
Next, we group the channels according to the related topics using LDA.
%We use the values obtained from LDA as input features to K-means~\cite{hartigan1979ak}, a classical clustering algorithm based on partitioning.
%\begin{figure}
%  \includegraphics[width=.5\textwidth]{images/violeted_terms.pdf}
%  \caption{Telegram service messages. %related to violation of terms of service.
%  }
%  \label{fig:violeted_terms}
%\end{figure}
Tab.~\ref{tab:english_topics} lists the topics discovered in the TGDataset and the number of channels for each. These emerged topics are quite different from those covered by the seed channels (see Sec.~\ref{sec:data_collection}).
Indeed, some disappeared (\eg Psychology, Marketing), while other interesting topics came up. 
\mmnote{
This shift in topic distribution is due to the snowball approach, which adds new channels linked by message forwarding to the dataset. As a result, channels from larger, more interconnected networks with higher forwarding activity are more likely to be included. This method naturally shifts the dataset toward more viral or widely shared topics.}

 %, Economics, Sales, Video Movies, Business Startups, Medicine, Psychology, and Fashion \& Beauty disappeared, 
 
Among the new topics, one is related to carding, the practice of selling full details of stolen credit cards or selling prepaid cards or other goods purchased with them. 
Similarly to what happens in dark web forums~\cite{kigerl2020behind}, carders (the people who own the stolen credit cards) use Telegram channels to place gift cards or goods for sale. In the TGDataset, we find 1,489 channels (7.53\% of the English part of the dataset) offering this service. In particular, Telegram marked 45 carding channels as scams, suggesting that some of them do not deliver the service they offer. 

Another unusual cluster containing 989 channels (5\% of English channels) is the one about extremists and radicals. Here, several channels promote white supremacy, as well as other Nazi ideologies and conspiracies against the white race (the title of one of these channels is \textit{White Genocide Immigration Anti-White Agenda}). Interestingly, there are four channels in this category having the verified status. 
One channel (\textit{"ISIS Watch"}) just publishes daily updates on banned terrorist content.
Instead, the other three channels are related to official news agencies of the Russian government and are no longer accessible on Telegram. Moreover, some of their messages we collected are obscured by the platform. 
%with the service message \textit{"This channel cannot be displayed because it violated local laws."}. 
To investigate this aspect deeper, we analyze the messages obscured by Telegram as they incited violence, published illegal pornographic content, or shared content protected by copyright. 
In this case, Telegram replaced some or all of the channel's messages with text explaining the reasons for obfuscation. 
Interestingly, the channel itself and the metadata (\eg the posting date) related to the original messages are still available. 
\begin{table*}[h!]
    \centering
    \small
    \begin{tabular}{l|cc}
    \toprule
        Message description & Number of messages & Number of channels\\
        \midrule
        This channel can’t be displayed because it violated Telegram's Terms of Service. & 428,793 & 2,546\\
        This channel can’t be displayed because it was used to spread pornographic content. & 263,643  & 969\\
        This channel can’t be displayed because it violated local laws. & 145,448 & 847\\
        This channel is unavailable due to copyright infringement. & 48,402 & 1,633\\
    \bottomrule
    \end{tabular}
    \caption{Number of posts obscured by each service message and number of channels presenting a specific service message.}
    \label{tab:violated_terms_messages} 
\end{table*}
As shown in Tab.~\ref{tab:violated_terms_messages}, the service message most used by Telegram to obscure a post is the one about the violation of Telegram's Terms of Service (428,793 messages). Follow the one about the spread of pornographic content (263,643 messages), the violation of local laws (145,448 messages), and copyright infringement (48,402 messages).
Finally, the other topics are more aligned with the ones covered by the seed channel of the TGDataset.

\begin{table}
    \centering
    \small
    \caption{Number of channels, scams, and verified ones belonging to each discovered topic.}
    \label{tab:english_topics}
    \begin{tabular}{l|c|c|c}
    \toprule
        Topic & \# channels & \# scam & \# verified\\ 
        \midrule
        Religion & 4,725 (23.90\%)           & 0  & 5 \\
        US news & 2,948 (14.91\%)            & 38 & 51\\
        Video-game modding & 1,957 (9.90\%) & 14 & 0 \\
        Covid & 1,716 (8.68\%)              & 0  & 5 \\
        Carding & 1,489 (7.53\%)            & 45 & 0 \\
        Entertainment & 1,440 (7.28\%)      & 0  & 3 \\
        World news & 995 (5.03\%)           & 0  & 17\\
        Extremists and radicals & 989 (5.00\%)       & 0  & 4 \\
        Indian education & 939 (4.75\%)     & 1  & 6 \\ 
        Software & 871 (4.41\%)             & 2  & 14\\
        Porn & 830 (4.20\%)                 & 5  & 0 \\
        Crypto & 563 (2.85\%)               & 5  & 10\\
        Social & 306 (1.55\%)               & 0  & 4 \\
        \bottomrule
    \end{tabular}
\end{table}

\subsection{Temporal aspect}

\label{sec:temporal}
\begin{figure}
  \Description{A time series plot showing the creation dates of Telegram channels over time, broken down into Russian, English, and all channels. The plot includes two significant events marked with dots: a green dot indicating when Russia banned Telegram, and an azure dot marking WhatsApp's privacy policy announcement.}
  \includegraphics[width=.49\textwidth]{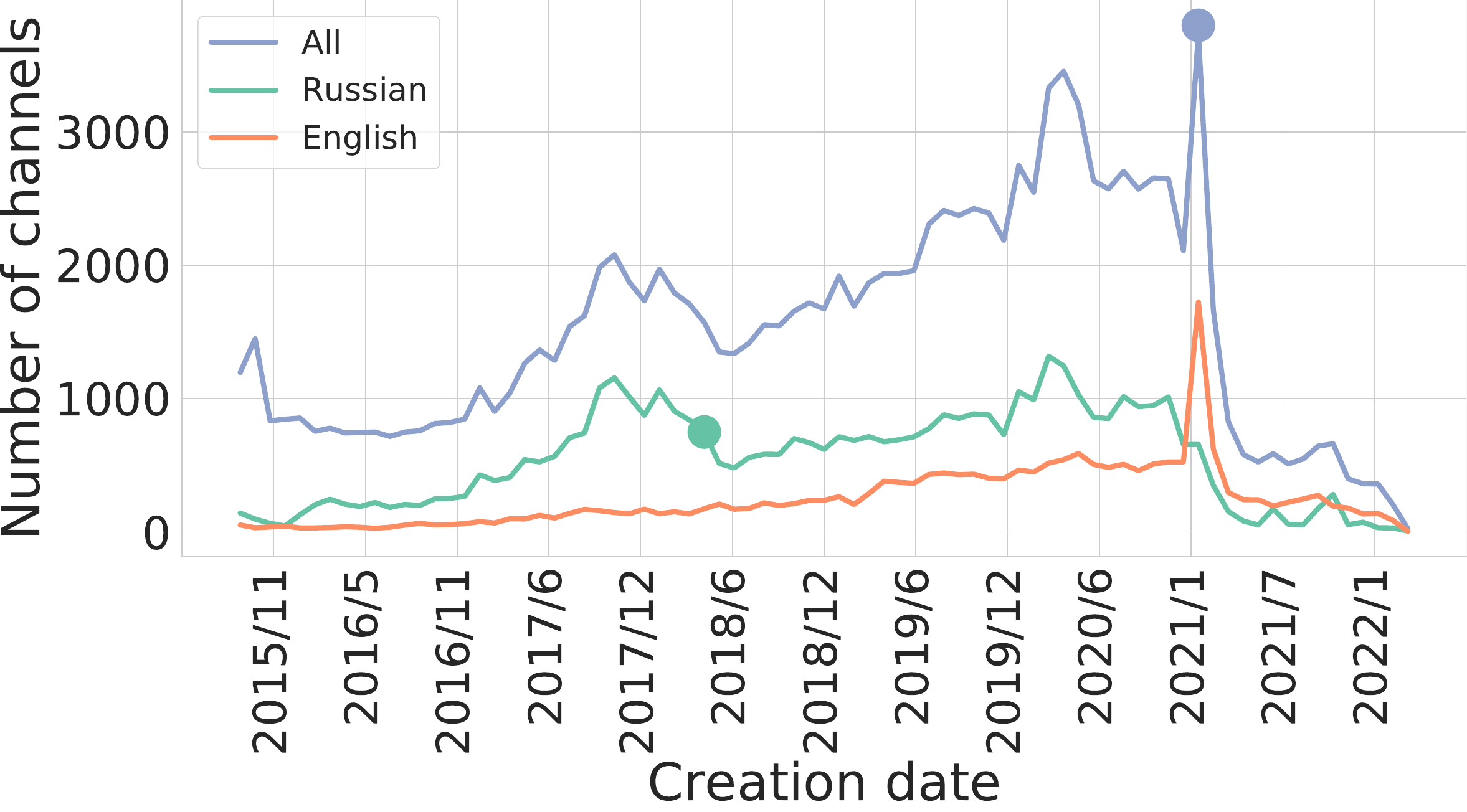}
  \caption{Creation date of Russian, English, and all the channels. The green dot represents the date when Russia banned Telegram, while the azure one indicates WhatsApp's announcement about privacy policy.}
  \label{fig:creation_date}
\end{figure}

Another interesting aspect is the creation date of the collected channels since, from this information, it is possible to understand how Telegram evolved through time. 
Fig.~\ref{fig:creation_date} shows the monthly creation rate of the channels contained in the TGDataset. In particular, the blue line represents the overall number of channels created on a certain date, while the green and the orange lines the number of Russian and English channels, respectively. 
As we can see, the number of channels created is steadily increasing, indicating that the Telegram platform is becoming more popular daily. Indeed, also many governments choose to use Telegram channels to communicate information regarding COVID-19~\cite{covidTelegram}. 
Until the beginning of 2018, the growth of channels is led by Russian-speaking users. However, in the first months of 2018, it is possible to note a sudden drop in daily created channels, likely due to the Russian ban of Telegram (green dot in Fig.~\ref{fig:creation_date}). Nonetheless, Russian channels, even if at a lower rate, continue to appear every day. Finally, in the first months of 2021, there was an abrupt increase of English-language spoken channels that overtook the Russian ones for the first time in Telegram history, according to our data. 
This event coincides with the change of WhatsApp's privacy policy (azure dot in Fig.~\ref{fig:creation_date}) and the migration of many users from WhatsApp to Telegram~\cite{whatsappTelegram}.

\section{Use cases}
The TGDataset has already been utilized in various studies~\cite{la2023sa, la2024pretending, imperati2023conspiracy}. For instance, La Morgia et al.~\cite{la2023sa} analyzed the problem of fake accounts on Telegram using the TGDataset, while Imperati et al.~\cite{imperati2023conspiracy} employed the TGDataset to identify 17,829 channels disseminating conspiracy-related content. 
In the following, this section describes other scenarios that can be explored and analyzed using the TGDataset.

\textbf{Political leaning and presence of questionable contents. }
Previous studies show the presence of moderation in OSNs (\eg Twitter) produces a significant reduction of questionable content, with respect to not moderated OSNs (\eg Gab)~\cite{etta1}, and that the information diffusion is biased by the political leaning of the community~\cite{cinelli2021echo}. %These studies quantify very well these phenomena in OSNs (\eg Twitter, Gab, Facebook, Reddit). 
However, very little is known about the loosely moderated Telegram platform. Thus, we use the TGDatset to perform the first high-level investigation on Telegram.
To this end, as done in~\cite{etta1,etta2022covid}, we leverage Media Bias/Fact Check (MBFC)~\cite{mediabiasfactcheck}, an independent fact-checking organization that rates the bias of news outlets. In particular, MBFC classifies the political bias of news outlets with one of the following labels: \textit{Extreme Left, Left, Left-Center, Least-Biased, Right-Center, Right,
Extreme Right}. Additionally, MBFC provides a second label to quantify the reliability of the outlet, categorizing outlets as \textit{Reliable} and \textit{Questionable}. 

For each English channel in the TGDataset, we label the links it contains according to the Media Bias/Fact Check (MBFC) categorization. We then assigned each channel the most frequently chosen label among its categorized links. This approach classifies 9,045 channels out of 19,768; the remaining channels did not share news or contain sources not present in MBFC.
Upon aggregating channels into left-leaning (Left-Center, Left, Extreme Left) and right-leaning (Extreme Right, Right, Right-Center) groups, we found a nearly balanced distribution: 4,491 channels shared right-leaning news, and 3,900 shared left-leaning news. However, 2,748 channels were identified as spreading extreme right content, while none were found to be spreading extreme left content.
Among the top four topics with the most categorized channels (World News, US News, Religion, and Covid), World News is the most left-leaning, with over 60\% of the channels associated with left-leaning outlets. Channels on the topic of Religion were almost evenly split between left and right. Conversely, channels dealing with US News and Covid were predominantly right-leaning. Specifically, the extreme right was underrepresented in World News (76 channels) but predominant in Religion (590 channels), US News (1,144 channels), and Covid (601 channels).
These three topics—Religion, US News, and Covid—also had the highest number of channels sharing content from questionable sources, with 1,353, 2,320, and 1,277 channels, respectively. Analyzing the reasons behind their questionable classification, we find that 55\% of Religion, 80\% of US News, and 85\% of Covid questionable channels shared content from outlets associated with conspiracy theories by MBFC.

This analysis reveals a significant presence of Telegram channels disseminating news from unreliable sources. Thus, the TGDataset could be a valuable resource for researchers aiming to understand the spread of misinformation, fake news, and conspiracy theories.
\mmnote{This use case involves using the content of the text messages posted by the channels (represented by the \textit{Message} field) to extract links for further processing, and the forwarding information, including whether the message was forwarded and its source (contained in the \textit{Is Forwarded} and \textit{Forwarded From} fields).}

\textbf{Study of channels spreading conspiracy theories. } 
\label{sec:sabmyk}
In the previous use case, we discovered that there are channels that spread news related to conspiracy theories. Here, we attempt to identify them through community detection.
A community in a graph is a subset of nodes that are densely connected to each other and weakly connected to nodes in other communities.

For this study, we represent our dataset as a directed graph $G=(V, E)$ in which nodes in $V$ are the channels, and edge $u \rightarrow v$ in $E$ represents the presence in channel~$u$ of a message originally posted in $v$ and forwarded to~$u$ by the admin of channel~$u$. Since the users of channel~$u$ can navigate the forwarded message and land on channel~$v$, the edge represents in a natural way the possible flow through channels of users following forwarded messages. To build and analyze our graph, we use the NetworkX library~\cite{hagberg2008exploring}.

Then, we use the Leiden algorithm~\cite{traag2019louvain}, an algorithm for community detection that improves the Louvain algorithm~\cite{ghosh2018distributed}. 
\mmnote{To ensure the accuracy of our partition, we used modularity, a validation metric that measures how much better our partitioning is compared to a random partition. Thus, we compute the optimal number of communities with respect to modularity, achieving a high score of 0.78, which indicates a strong community structure remarkably better than random partitioning.}
This approach finds a partition with 311 communities, 47 of which with more than one node.
One of the communities, which consists of 236 channels and where English is the common language, has the peculiarity that all posts and messages forwarded from one channel to the other are about a new conspiracy theory called Sabmyk. Sabmik is a conspiracy theory that proposes itself as a better alternative to QAnon. It promotes a quasi-religion centered around a messianic figure known as Sabmyk~\cite{independetSabmyk}. 
This community of channels has, collectively, more than 1 million subscribers. The most popular channel in the community is \textit{Great Awakening Channel} with 119,103 subscribers. According to our data, most of these channels were created at the beginning of 2021---130 of them (55.08\%) between January and February---with the latest in our dataset created in February 2022. The nature of these channels is very different: Some of them are fake channels of celebrities (\eg Mel Gibson, Keanu Reeves, Kanye West), while others target news outlets or official channels of national bodies (\eg U. S. Marines Channel, U. S. Navy Channel) and others conspiracy theories like QAnon. Regardless of the name of the channel in the community, the content is always the same. Indeed, they recycle and forward messages among them: Out of the 1,203,986 messages published by these 236 channels, only 65,602 are unique.
With this analysis we scratch the surface of the Sabmyk network. Still, more investigation are needed to understand how Sabmyk lures such a large number of subscribers, its ultimate goal, and its connection to other conspiracy theories. 

\mmnote{Performing this study requires several key elements of the dataset. 
By analyzing the content of the \textit{Message} field, it is possible to identify the topics within conspiracy theories.
With the \textit{Is Forwarded} field, we can understand if the message has been forwarded, and in the case it is, which was the source channel leveraging the \textit{Forwarded From} field. This information makes it possible to build a graph representing connections between channels and to detect the communities as we did for our analysis.}
%The \textit{Author} field is necessary for determining the original creator of each message, which helps to trace the source of the forwarded content. The \textit{Is Forwarded} field identifies whether a message has been forwarded, which is crucial for constructing a graph representing connections between channels. Lastly, the \textit{Forwarded From} field is used to identify the source channel from which a message was forwarded, forming the directed edges of the graph required for community detection.}

\textbf{Carding and underground markets.} In Section~\ref{sec:topic_modeling}, we examine the main topics within our dataset. Upon investigating the channels associated with these topics, we discovered several engaging in borderline activities, such as running underground marketplaces. These channels sell electronics, Netflix accounts, and hacking tools for low prices. These types of marketplaces are often associated with carding.
Consequently, it is worth delving further into these Telegram channels to determine the types of goods being sold, their origins, whether the channels are just an attempt to fraud subscribers or if they are actually engaged in more illegal activities, and whether there is any correlation with Dark Web marketplaces.

\mmnote{For this analysis, the \textit{Message} field is crucial for identifying the types of goods sold within the channels, such as electronics, Netflix accounts, and hacking tools. Analyzing the content of the messages and extracting the URLs makes it possible to determine whether the channels are involved in fraudulent activities or illegal transactions.}

\textbf{Copyright Infringement and personal content.}
Recently, Telegram hit the news several times for its use for the distribution of copyrighted content (\eg movies and software).
For instance, Italian authorities seized 545 channels for copyright infringement~\cite{teleitalianchannel}. However, the issue extends beyond copyright violations. Indeed, the platform is frequently misused to distribute personal content without consent, such as the unauthorized distribution and sale of an Indian teacher's course material~\cite{teleindianteacher}. Even more troubling, Telegram is also used to share leaked nude photos~\cite{teleLeakNudes} or facilitate revenge porn~\cite{teleRevengePorn}.
Joining the information about the topic of the channels, and the description of the messages
removed by the platform (Tab.~\ref{tab:violated_terms_messages}), we can observe that the four categories in which is most present the diffusion of copyrighted material are: Carding (448 channels), Indian Edu (219 channels), Video-game modding (208 channels), Adult content (157 channels). 
Starting from these channels, it is possible to study the diffusion of the phenomenon, understand how these channels are organized, and how they monetize.
\mmnote{For this analysis, the \textit{Message} field is essential for identifying illicit content, such as copyrighted material, while the \textit{Description} field could provide additional context about channel activities.}

\textbf{Cryptocurrencies frauds.}
We also found several (563) channels related to the cryptocurrency world. It is well known that Telegram channels are exploited by Pump and Dump groups~\cite{xu2019anatomy,morgia2021doge} that perpetrate frauds on the crypto-market. We notice that some of the Pump and Dump channels monitored in~\cite{la2020pump} are also in our dataset. Previous works monitored the channels vertically, focusing on the fraud and the groups' mechanics. Thus, \mmnote{using the \textit{Message}, \textit{Forwarded from}, and \textit{Description} fields}, the TGDatast can help understand how these channels are connected, if they operate together, how they promote their services, and likely discover new channels that carry out similar activities.

\textbf{Violence and extremism.}
According to several newspapers, the Capitol Hill riot of January 2021 was planned months before also leveraging Telegram channels and groups~\cite{capitol_riot_on_telegram_forbes, capitol_riot_on_telegram_vox}. In response to this event, the platform has intensified its monitoring and obscured dozens of public channels promoting calls to violence~\cite{telegramBanCapitolRiot}.
Nonetheless, despite the commitment of Telegram to obfuscate these channels, within our dataset, there are several public channels promoting the spread of neo-Nazi ideologies or calls to violence (\eg \textit{White Aryan Woman}, \textit{Feuerkrieg Division **OFFICIAL**}).
Therefore, the problem is still far from being solved.
The TGDataset can help study and characterize these channels. Indeed, the main problem with these channels is the absence of a dataset containing them, as they are obscured or blocked once discovered. Instead, the TGDataset includes several unblurred channels of this typology that can be used to analyze their features and build a machine-learning model able to detect them automatically.
\mmnote{This study requires the \textit{Message} and \textit{Description} fields to analyze harmful content and obtain context on the purposes of these channels, and the \textit{Forwarded from} field to trace the channel behind it.}

\mmnote{
\textbf{Temporal analyses and information propagation.} The TGDataset represents a valuable resource for exploring the temporal dynamics of information spread by leveraging message timestamps and channels' creation dates. Message timestamps (\textit{Date} field in Text messages and Media messages) facilitate granular analyses of dissemination patterns, enabling researchers to track how content spreads within and across channels. Instead, channels' creation dates (\textit{Creation date} field) enable the analysis of the correlation between a channel's longevity and its influence on information propagation.
These temporal attributes support a variety of studies. For example, if combined with the forwarding information (\textit{Is forwarded} and \textit{Forwarded from} fields in Text messages and Media messages), they allow for tracing the rise and evolution of trending topics, assessing the velocity of information diffusion for different content types, identifying peak activity periods for specific topics or channels, and examining shifts in information flow patterns over time.
Moreover, beyond individual channels, the communities identified in the TGDataset present opportunities for studying inter-community information diffusion. Investigating these larger network structures could provide valuable insights into how information propagates across interconnected groups within Telegram, unveiling the mechanisms that drive broader dissemination trends.
}

\section{Ethical Considerations}
\label{sec:ethical_considerations}
This paper presents the TGDataset, a new dataset that includes 120,979 Telegram channels and over 400 million messages. The data collection was conducted carefully to only include information from Telegram channels and exclude personal data such as usernames, phone numbers, and subscribed channels.

\mmnote{Our data collection process complies with Telegram’s Terms of Service~\cite{TelegramToS} and Telegram API's Terms of Service~\cite{TelegramAPIToS}, as there are no clauses that prohibit the collection of public chat data. Additionally, we have taken care to avoid flooding the platform with requests during data collection.}
\mmnote{
Moreover, our research complies with the academic research exemptions outlined in Article 85 of the GDPR~\cite{gdpr85}, which provide certain flexibilities for processing publicly available data in the interest of freedom of expression and academic purposes. As our work exclusively involves publicly accessible content and does not process private user data, it falls under the legitimate or public interest exemptions.
Furthermore, we adhere to the principle of data minimization, as emphasized in the GDPR's guidelines~\cite{gdpr_minimization}: We limit our analysis to admin-generated public posts, ensuring that only the information strictly necessary for our research objectives is processed.} 
 
Lastly, since the dataset includes channels that discuss controversial topics, some controversial messages may be present in the dataset. To comply with ethical standards, we did not download any images or include links in the public dataset as they may contain adult content or copyrighted material. Hence, according to our IRB's policy, we did not require explicit authorization to conduct our experiments.

\section{Limitation}
To create the TGDataset, we began with 180 seed channels and then employed a snowball technique to grow the dataset. This approach is limited to only reaching channels that are linked by message forwarding. Despite this constraint, we were able to uncover over 100,000 channels and could potentially discover even more through continued iterations. Nonetheless, there may be groups of channels that remain inaccessible from our seed channels.

In the TGDataset, channels are represented as a snapshot of their current state. Although some historical details about the channels are available in the dataset, such as their creation date and timestamp of messages, other information is not present. 
For instance, a channel during time can change its name, description, and of course, the number of subscribers. This information could be useful for reconstructing the evolution of the channels.
In the current release of the TGDataset, we do not retrieve the comments posted by users on channels' messages or the reaction emojis they may have used (such as the 'like' button on Facebook). While we have observed that many Telegram channels do not use these features, they can still offer valuable insights for future analyses.

\section{Conclusions  and Future works}
Telegram has gained significant popularity in recent years. As a result, it is crucial to study and understand the activity taking place on the platform. 

In this work, we present the TGDataset~\cite{TGDatasetRepo}, a collection of more than 120,000 public Telegram channels that, to the best of our knowledge, is the largest collection of channels publicly available.
After characterizing the main quantitative aspect of the TGDataset, we performed language detection to understand which are the most popular languages on the dataset. Then, we investigate the main topics covered by the English channels. 
Additionally, we publicly released the script we used to analyze it and the labeling we obtained (language and topic of the channels).
In this paper, we investigate a few possible use cases in which our dataset can be extremely useful. In particular, our preliminary study of the TGDataset revealed some Telegram channels spread questionable content and conspiracy theories. 
Moreover, we observed the presence of several borderline activities and channels dealing with dubious ethical content.
With the release of the TGDataset, we aim to provide a valuable resource to researchers that enables further investigation into these areas, leading to a more refined understanding of Telegram and helping to mitigate potential risks to users.

As future work, we plan to continue running our data collector and further enlarge the TGDataset, releasing new versions of the dataset at regular intervals of time. Moreover, we intend to overcome the actual limitations by recording channel updates (channel's name, subscribers, and description), adding to the list of channels to monitor also those referenced by link, inserting the possibility to add new seed channels, and collecting replies to messages.

\section{Acknowledgments}
This work has been partially funded by projects: MUR National Recovery and Resilience Plan, SERICS (PE00000014); and ST3P (B83C24003210001) under the "Young Researchers 2024-SoE" Program funded by the Italian Ministry of University and Research (MUR).

\bibliographystyle{ACM-Reference-Format}
\bibliography{sample-base}

\end{document}